\documentclass{article}
\usepackage{amssymb,amsmath}
\usepackage{graphicx,rotate}
\usepackage{cite}
\usepackage{Arrayeq}

\usepackage{color}

\def \order(#1){{\cal O} \left(#1 \right)}

\evensidemargin=\oddsidemargin
\def\Eqn#1{Eq.\ (\ref{#1})}

\begin{document}


\begin{flushright}
SINP/TNP/2012/16
\end{flushright}

\begin{center}
{\Large \bf Modified Higgs couplings and unitarity violation } \\
\vspace*{1cm}  {\sf
    Gautam Bhattacharyya, Dipankar Das, Palash B. Pal}
\\
\vspace{10pt} {\small } {\em Saha Institute of Nuclear
    Physics, 1/AF Bidhan Nagar, Kolkata 700064, India}

\normalsize
\end{center}

\begin{abstract}

Prompted by the recent observation of a Higgs-like particle at the
CERN Large Hadron Collider (LHC), we investigate a quantitative
correlation between possible departures of the gauge and Yukawa
couplings of this particle from their Standard Model expectations and
the scale of unitarity violation in the processes $WW \to WW$ and
$t\bar t \to WW$.

\end{abstract}

\bigskip

One of the crucial arguments for the existence of the Higgs boson in
the Standard Model (SM) is that, without it, the longitudinal vector
boson ($V_L$, where $V=W,Z$) scattering amplitudes at the tree level
would uncontrollably grow with the center of mass energy ($E$).  This
will result in the violation of `unitarity', thus implying breakdown
of quantum mechanical sense of probability conservation in scattering
amplitudes.  In the SM, the Higgs boson possesses appropriate gauge
couplings to ensure exact cancellation of the residual $E^2$ growth in
the $V_L V_L \to V_L V_L$ scattering amplitude that survives after
adding the gauge boson contributions.  It has been explicitly shown in
\cite{Lee:1977eg} how, for $E \gg M_V$, the $E^2$ dependence is traded
in favor of the unknown $m_h^2$, where $m_h$ is the Higgs boson mass.
From this it was concluded that $m_h$ should be less than about a TeV
for unitarity not to be violated.  An intimate relationship between
unitarity and renormalizability adds a special relevance to this
issue.  For a renormalizable theory the tree level amplitude for $2
\to 2$ scattering should not contain any term which grows with energy
\cite{Cornwall:1974km}.  In perturbative expansion of scattering
amplitudes these energy growths must be canceled order by order
\cite{Chanowitz:1978mv}.  It has been shown that the energy dependent
terms in tree level amplitudes get exactly canceled if the couplings
satisfy certain sets of `unitarity sum rules' \cite{Gunion:1990kf}.
It has also been realized that the presence of the Higgs boson is not
the only option to satisfy these sum rules \cite{Csaki:2003dt,
  Lahiri:2011ic}.

Meanwhile, a Higgs-like particle has been observed with a mass of
around 125 GeV by the ATLAS and CMS collaborations of the LHC
\cite{:2012gk, :2012gu}.  This is much below the upper limit coming
from unitarity violation mentioned above.  If this particle indeed
turns out to be the SM Higgs, then the scattering amplitudes involving
not only the longitudinal vector bosons but any other SM particles as
external states would be well behaved for arbitrarily high energies.
However, the recent observation of some excess events in the $h \to
\gamma\gamma$ channel, as well as large errors associated with other
decay channels, has fuelled speculation that Higgs couplings to
fermions and/or gauge bosons might not be exactly as predicted by the
SM \cite{Carena:2012xa}.  There are more than one ways to modify the
Higgs couplings. One way is to hypothesize that the $WWh$ and the
$ZZh$ couplings are modified; more specifically, enhanced with respect
to their SM values.  This would result not only in an increase in the
Higgs production cross section via vector boson fusion and associated
production, but also in an enhancement of the $W$-loop contribution to
$h \to \gamma\gamma$ decay.  But this would at the same time lead to
excess events in the $h \to WW^*$ and $h \to ZZ^*$ channels, something
which is not obvious from data.  It would also result in the violation
of unitarity in longitudinal gauge boson scattering channels. This was
indeed explored long back \cite{Cheung:2008zh}, however, in the
absence of the LHC data there was no motivation to study the
correlation between unitarity violation and the Higgs decay branching
ratios at that time.  If we refrain from adding any extra particle to
the SM and yet attempt to account for the excess in the diphoton
channel, the next natural choice would be to modify the Yukawa
coupling of the top quark.  As is already known, if we put the sign of
the top Yukawa coupling opposite to what it is in the SM, the $h \to
\gamma\gamma$ rate gets enhanced due to a constructive interference
between the $W$-loop diagram and the top-loop diagram
\cite{Carena:2012xa}.  One of the fall-outs of this sign flip is that
$t\bar t \to V_L V_L$ scattering no longer remains unitary.  In fact,
as we shall show, any non-trivial admixture of CP-even and CP-odd
states in the composition of the scalar particle jeopardizes the good
high energy behavior of the $t\bar t \to V_L V_L$ amplitude even if we
keep the moduli of the top Yukawa coupling and the Higgs gauge
coupling to their SM values.  The purpose of this paper is to
explicitly demonstrate how the scales of unitarity violation in $W_L
W_L \to W_L W_L$ and $t \bar t \to W_L W_L$ scattering processes
depend on the modification parameters of the gauge and the top Yukawa
couplings of the Higgs.  We demonstrate what an enhanced diphoton rate
may imply in this context.

In our analysis, we modify only the top Yukawa coupling, since the
other Yukawa couplings are numerically much less relevant.  We take
\begin{subequations}
\label{xfdef}
\begin{eqnarray}
g_{tth} &=& 
(1-f) (\cos \delta - i \sin \delta \gamma_5) \, g_{tth}^{\rm SM} 
= (1-f) e^{- i\delta \gamma_5} g_{tth}^{\rm SM} \,.
\label{fdef}
\end{eqnarray}
The parameter $f$ is a measure of the overall coupling of the Higgs
boson to the top quark, whereas $\delta$ is a parameter that
quantifies the mixture of CP-even and CP-odd components in the Higgs
boson.  We also modify the gauge couplings of the Higgs boson as
\begin{eqnarray}
g_{VVh} &=& (1-x) \, g_{VVh}^{\rm SM} \,,
\label{xdef}
\end{eqnarray}
\end{subequations}
where $V$ can be $W$ or $Z$, as said before.  We maintain equality
between the $WWh$ and $ZZh$ couplings to respect custodial symmetry.
The parameters $x$, $f$ and $\delta$ are all real, and they all vanish
in the SM.

We now comment on the existing experimental constraints on these
modification parameters.  First, it has been shown in
\cite{Azatov:2012bz} that precision electroweak measurements imply
$-0.2 \le x \le 0.1$ at 95\% C.L.\ for $m_h = 125$ GeV and $m_t = 173$
GeV, while from the recent LHC Higgs data analysis the 95\%
C.L.\ range has been estimated to be $-0.4 \le x \le 0.4$
\cite{Plehn:2012iz, Giardino:2012dp}.  Second, the allowed range of
$f$ can be extracted from recent fits of modified Higgs couplings
against the LHC data.  For example, for $x=0$, the range is $-0.1 < f
< 0.6$ for values of $\delta$ fixed at $0$ and
$\pi$~\cite{Giardino:2012dp, Espinosa:2012im}. Note that similar
bounds have been obtained by the authors of
Ref.~\cite{Banerjee:2012xc}, who considered a phase in the effective
coupling due to an absorptive part in the amplitude.  In this paper,
we take a more conservative approach and consider a hermitian Yukawa
Lagrangian.

\begin{figure}
\null\hfill 
\includegraphics[width = 0.39\textwidth]{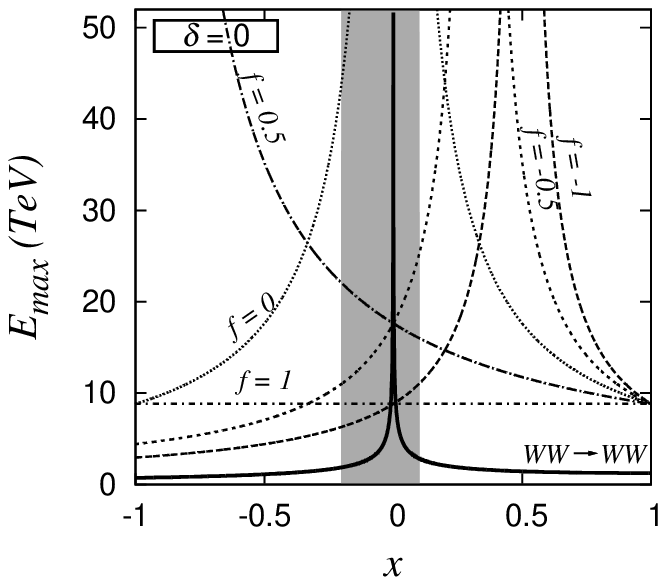} \hfill
 \includegraphics[width = 0.39\textwidth]{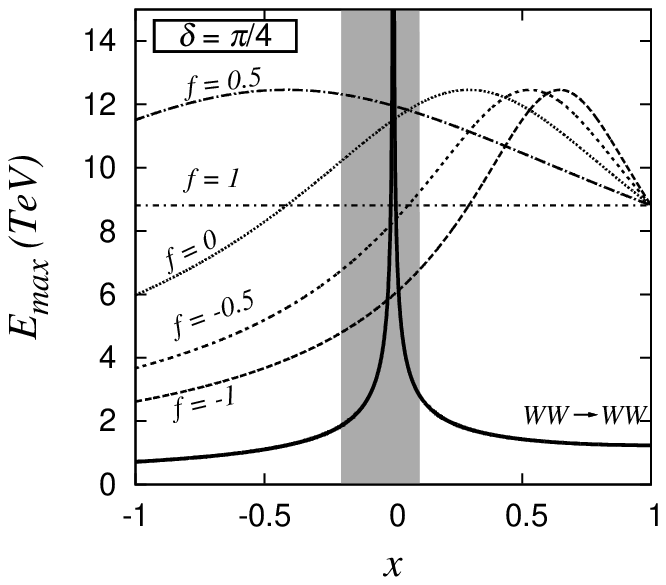} 
\hfill\null\\
\null\hfill 
 \includegraphics[width = 0.39\textwidth]{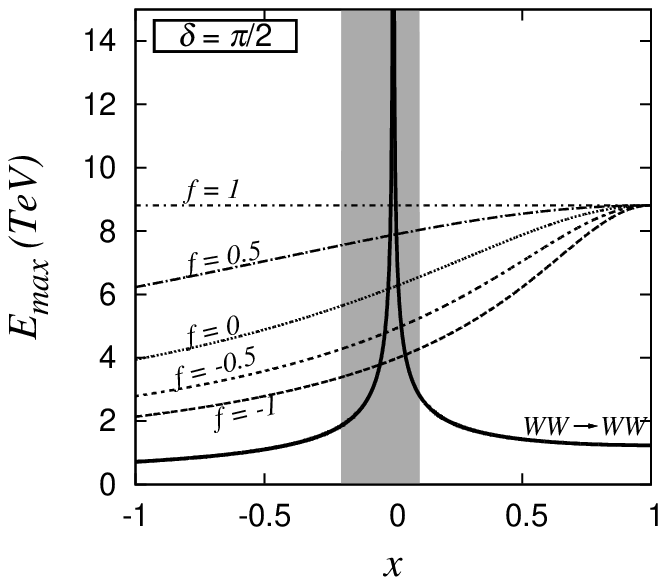} \hfill
 \includegraphics[width = 0.39\textwidth]{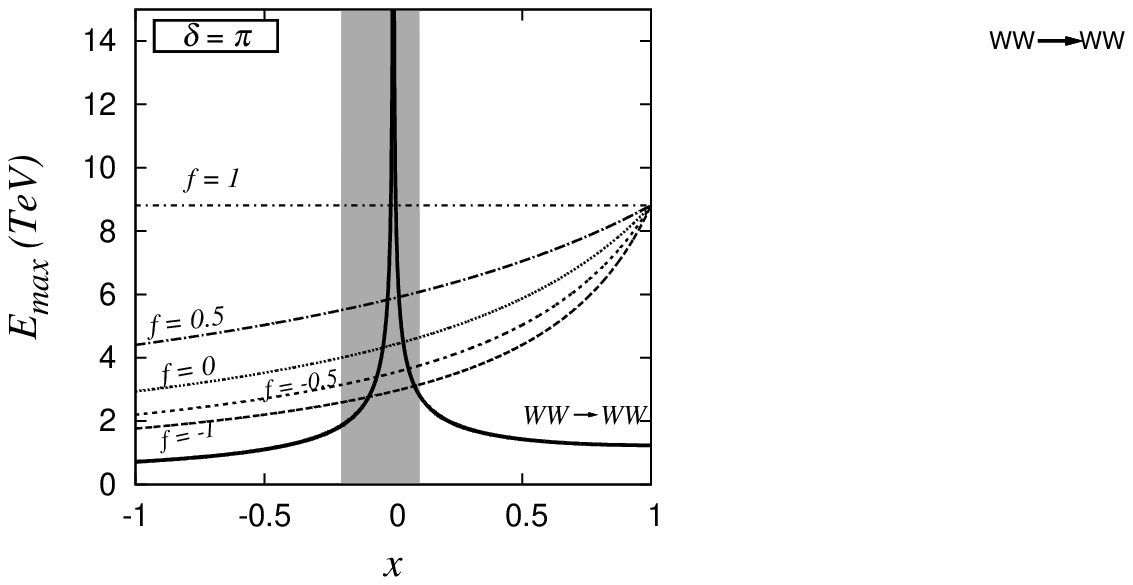}
\hfill\null

 \caption{\em Unitarity violation scale as a function of $x$, for
   specific values of $f$ and $\delta$.  For each panel, the scale
   coming from the elastic $WW\to WW$ scattering has been marked.  The
   other lines come from $t\bar t \to WW$ scattering for various
   values of $f$.  The vertical shaded region represents the range of
   $x$ consistent with electroweak precision data.  Note the different
   scale on the vertical axis for the plot with $\delta=0$.}
 \label{xEplot}
\end{figure}
With the modifications prescribed in \Eqn{xfdef}, one should examine
unitarity constraints on scattering processes involving the top quark
and the $W$-boson.  Note that we will talk about the longitudinally
polarized component of the $W$-boson only, dropping the polarization
subscript $L$ which is implicitly assumed.  We have looked at the
energy dependence of the elastic scattering $WW \to WW$ and the
inelastic scattering $t\bar t \to WW$.  The scattering amplitudes that
we find are as follows:
\begin{subequations}
\label{amplitudes}
\begin{eqnarray}
{\cal A}^{WW\to WW} &=& 2\sqrt{2}G_F E^2 (2x-x^2)(1+\cos\theta) +
\cdots \,, 
\label{ampWWWW}  \\ 
 {\cal A}^{t\bar t \to WW} &=& 
2\sqrt{2} G_F E m_t Y(x,f,\delta) + \cdots \,, 
\label{ampWWtt}
\end{eqnarray}
\end{subequations}
where the dots indicate sub-leading terms in energy which do not
concern us, $\theta$ is the scattering angle, and 
\begin{eqnarray}
Y(x,f,\delta) = \mp \Big[ 1 - (1-x)(1-f)e^{\mp i\delta} \Big] \,,
\label{Y}
\end{eqnarray}
where different signs correspond to different combinations of
helicities \cite{Whisnant:1994fh}.  The scattering amplitude can be
expanded in terms of partial waves \cite{Lee:1977eg}:
\begin{eqnarray}
 {\cal A}(\theta)=16\pi\sum_{l=0}^{\infty}(2l+1)a_lP_l(\cos\theta) \, .  
\end{eqnarray}
The unitarity condition $|a_0|\le 1$ puts upper limits on the center
of mass energy in each of these processes.  These limits are as follows:
\begin{subequations}
\label{u12}
\begin{equationarray}{rcll}
E \leq E_{\rm max}^{WW} & = & \left(\frac{4\sqrt{2}\pi}{G_F} \;
\frac{1}{|2x-x^2|}\right)^\frac{1}{2} & \text{[from $WW \to
     WW$]} \, ;\label{u1} \\ 
E \leq E_{\rm max}^{tt} & = & \frac{4\sqrt{2}\pi}{G_Fm_t} \;
\frac{1}{|Y(x,f,\delta)|} \hspace{20mm} &  \text{[from $t\bar
    t \to WW$]} \, .\label{u2} 
\end{equationarray}
\end{subequations}
Because only $\cos\delta$ appears in $|Y|$, we can take $\delta$ in
the range $[0,\pi]$.  Without any loss of generality, we can take
$1-f\geq 0$ to cover the entire parameter space.  In passing, let us
add that the constraints from $t \bar t \to ZZ$ is the same in the
leading order in $E$ as that given in \Eqn{u2}.

We now discuss the numerical dependence of the unitarity violation
scale on the nonstandard parameters expressed through our master
equations given in \Eqn{u12}.  Our results are displayed in
Fig.~\ref{xEplot}.  The different panels correspond to different
choices of $\delta$, as indicated in the figure.  For the $WW \to WW$
scattering amplitude which grows as $E^2$, there is contribution
coming from Higgs mediated diagram and therefore it depends on $x$,
but there is no dependence on $f$ and $\delta$ since the top-Higgs
coupling is not involved.  The latter coupling is of course relevant
for the $t\bar t \to WW$ scattering, and the Higgs mediated graph is
sensitive to all the three nonstandard parameters, i.e.  $x$, $f$ and
$\delta$.  In all the panels the lines titled $WW \to WW$, obtained by
plotting \Eqn{u1}, show the scale of unitarity violation as the $WWh$
coupling departs from its SM value.  The other lines mark the
unitarity violation scale arising from $t \bar t \to WW$, and are
obtained from Eq.~(\ref{u2}).  In the limit $x=1$, i.e.  when the
Higgs either does not exist or does not couple to $W$, unitarity is
violated at a pretty low scale, $E_{\rm max}^{WW} \approx 1.3$ TeV.
As $x$ approaches zero, $E_{\rm max}^{WW}$ goes up.  On the other
hand, the limit $f=1$ implies that the Higgs does not couple to the
top quark, so in this limit the Higgs mediated graph for $t\bar t \to
WW$ would not exist, and hence, the unitarity violation scale arising
from the above scattering would be independent of $x$ and $\delta$.
Similar things happen in the limit $x=1$, causing the unitarity
violation scale from $t\bar t \to WW$ to be independent of $f$ and
$\delta$.  This is precisely the reason as to why the horizontal $f=1$
line in all the panels meet the curvy lines for other values of $f$ at
one single point which is at $x=1$ corresponding to $E_{\rm max}^{tt}
\approx 9$~TeV.

An important observation at this stage is the following: for
$\delta\neq0$ and $\delta\neq\pi$, the process $t\bar t \to WW$ is not
unitary regardless of the choice of $x$ and $f$.  The vertical shades
in the four panels restrict the values of $x$ within the zone allowed
by precision tests.  One thing is quite clear that if $x$ happens to
take a value near the edge of the shade in any panel, the unitarity
violation would set in for $WW \to WW$ at a scale much lower than
where it would happen for $t\bar t \to WW$, which is easily understood
from the $E^2$ versus $E$ growth in the two amplitudes.  But if $x$
settles at a much smaller value, as one can see from the different
panels, the unitarity violation scales from these two amplitudes get
closer and at some point the hierarchy mentioned earlier is reversed.

We now consider the decay of the 125\,GeV particle into two photons.
Two-photon final states have a definite CP property, more
specifically, a definite parity.  As a result, if the initial
spin-zero state is not an eigenstate of parity, the parity-even and
parity-odd components will contribute incoherently.  For the sake of
simplicity and to provide intuitive feel for easy comparison with
standard expressions, we consider the decay of a CP-even scalar state
only, which amounts to taking $\delta = 0$ or $\pi$.

The decay $h\to \gamma\gamma$ proceeds dominantly through a $W$ boson
loop and a top loop diagram.  For a CP-even $h$, the decay width is
given by \cite{Djouadi:2005gi}:
\begin{eqnarray}
 \Gamma (h\to \gamma\gamma) = \frac{\alpha^2g^2}{2^{10}\pi^3}
 \frac{m_h^3}{M_W^2} \Big|F_W + \frac{4}{3}F_t \Big|^2 \,.
\end{eqnarray}
For the SM, the values of $F_W$ and $F_t$ are given by
\begin{eqnarray}
 F^{\rm SM}_W = 2+3\tau_W+3\tau_W(2-\tau_W)f(\tau_W) \,,  \qquad 
 F^{\rm SM}_t = -2\tau_t \left[1+(1-\tau_t)f(\tau_t)\right] \,,
\end{eqnarray}
where
\begin{eqnarray}
\tau_x \equiv (2m_x/m_h)^2 \,.
\end{eqnarray}
For $m_h\approx125 ~\textrm{GeV}$, $\tau_x >1$ for both $x=W,t$.  In
this situation, 
\begin{eqnarray}
f(\tau) =
\left[\sin^{-1}\left(\sqrt{1/\tau}\right)\right]^2 \,.
\end{eqnarray}
Using the modified Higgs couplings of \Eqn{xfdef}, the expressions of
the $W$ and top loop contributions are obtained by replacing $F^{\rm
  SM}_W$ and $F^{\rm SM}_t$ by
\begin{eqnarray}
 F_W = (1-x) F^{\rm SM}_W \,, \qquad 
 F_t = (1-f)e^{-i\delta} F^{\rm SM}_t \,,
\end{eqnarray}
where $\delta$ is either zero or $\pi$, as mentioned earlier.

We now estimate how the Higgs production cross section would be
modified.  For 7(8)-TeV LHC, the top loop driven gluon-gluon fusion
channel contributes around 85\% of the total cross section, while the
associated production and the vector boson fusion together almost
account for the remaining 15\% \cite{Djouadi:2005gi}.  The production
cross section would then be modified roughly by the factor
\begin{eqnarray}
\frac{\sigma(pp\to h)}{\sigma^{\rm SM}(pp\to h)} 
= \frac{(1-f)^2\sigma_{\rm G}+(1-x)^2\sigma_{\rm V}}{\sigma_{\rm
    G}+\sigma_{\rm V}} \approx (1-f)^2 85\% + (1-x)^2 15\% \,.
\end{eqnarray}
As far as the different decay channels of the Higgs are concerned, for
$m_h \approx 125$ GeV, branching ratios of the SM Higgs boson are
roughly as follows: $58\%$ to $b\bar{b}$, $7\%$ to $\tau^+\tau^-$,
$3\%$ to $c\bar{c}$, $24\%$ to $VV^{*}$ and $8\%$ to $gg$
\cite{Djouadi:2005gi}.  We then express the modification of the total
decay width by the ratio:
\begin{eqnarray}
 \frac{\Gamma_h}{\Gamma^{\rm SM}_h} = (58\%+7\%+3\%) +
 (1-x)^2 24\% + (1-f)^2 8\% \, .
\end{eqnarray}
The above expressions lead us to define
\begin{eqnarray}
 \mu &=& \frac{\sigma(pp\to h)}{\sigma^{\rm SM}(pp\to h)} \cdot
 \frac{ \Gamma (h \to \gamma\gamma)}{ \Gamma^{\rm SM}(h
 \to \gamma\gamma)} \cdot \frac{\Gamma^{\rm SM}_h}{\Gamma_h}  \,.  
\label{mu}
\end{eqnarray}

\begin{figure}
\begin{minipage}{0.46\textwidth}
\centerline {\includegraphics{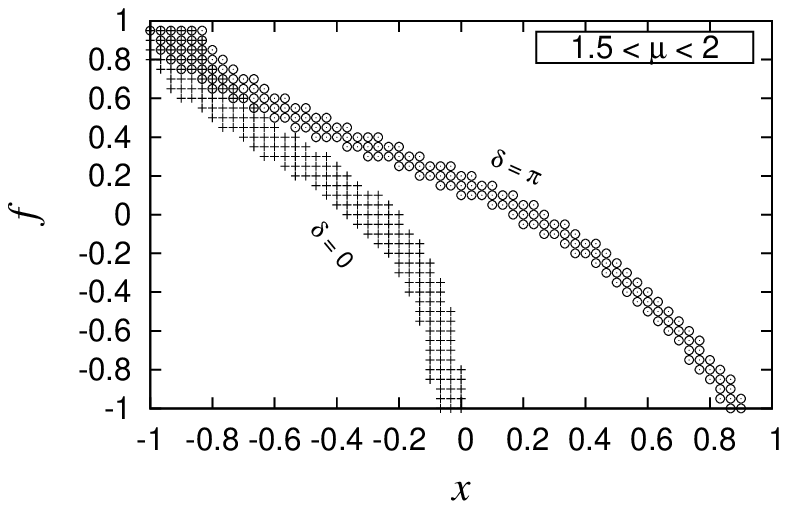}}
 \caption{\em Allowed regions in the $x$-$f$ plane that correspond to
   the diphoton enhancement ratio $\mu$ lying between 1.5 and 2, for
   $\delta=0$ and $\delta=\pi$.}
 \label{xf}
\end{minipage}
\hfill
\begin{minipage}{0.46\textwidth}
\centerline{\includegraphics{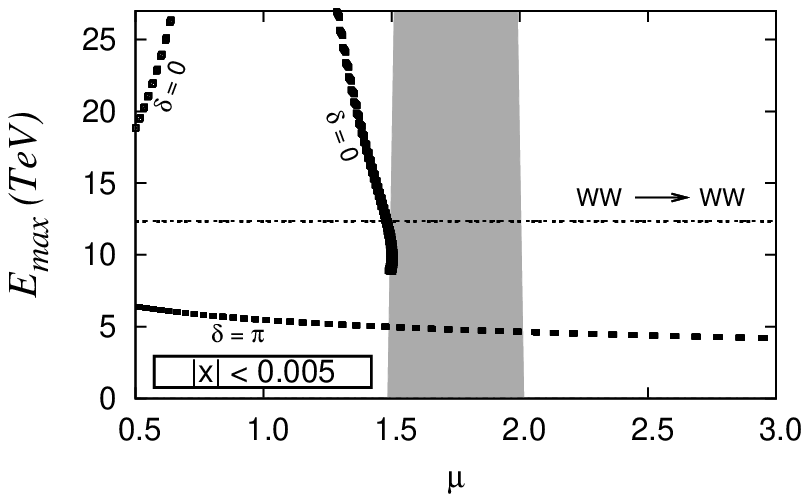}}
 \caption{\em Unitarity violation scale plotted against diphoton
   enhancement.  The vertical shaded region corresponds to $1.5 < \mu
   < 2.0$.}
 \label{emaxmu}
\end{minipage}
\end{figure}
In Fig.~\ref{xf} we have shaded different regions in the $x$-$f$
plane, for the two possible choices of $\delta$, which can account for
the apparent excess of the diphoton events.  Motivated by the recent
LHC data, we choose $\mu$ in the range $1.5$ to $2$ for the sake of
illustration.  For $x \approx0$ and $\delta=\pi$, we observe that 
\begin{eqnarray}
0.1 < f < 0.25 
\label{flimit}
\end{eqnarray}
which is roughly consistent with the limit quoted earlier in
connection with global fits.  Thus a {\em top-phobic Higgs}, which
corresponds to $f \to 1$, is highly unlikely.  We must admit though
that this comparison is not entirely fair as we have modified only the
top Yukawa coupling, while in the global fits all the Yukawa couplings
were modified.  We also admit that for the simplicity of illustration
we have not taken into account the efficiency factors in the
estimation of $\mu$.

In Fig.~\ref{emaxmu}, we have exhibited the correlation between the
unitarity violation scale and the diphoton enhancement ratio $\mu$.
For drawing this plot, we have varied $f$ between $-1$ and $+1$.
Keeping in mind the relative sensitivity of the two scattering
processes, we restrict $x$ in a rather narrow range: $-0.005 < x <
0.005$.  The horizontal line, appropriately labeled, corresponds to
the unitarity violation scale in $WW\to WW$ scattering with
$|x|=0.005$.  For smaller values of $x$, this line will appear at
higher energy.  The other curvy lines come from $t \bar t \to WW$ and
they correspond to two different choices of $\delta$, viz., zero and
$\pi$.  The thickness of the lines for $\delta=0$ and $\delta=\pi$
come from the range of $x$ just mentioned.  For $\delta=0$, it is hard
to achieve a value of $\mu$ as large as 1.5.  For $\delta=\pi$, it is
possible to obtain a value of $\mu$ in the range 1.5 to 2, as can be
seen by the corresponding line going through the vertical shade.  The
corresponding range of $f$, which can be read from Fig.~\ref{xf}, has
been mentioned in \Eqn{flimit}. It is worth noting from this figure
that for $\delta=\pi$, which facilitates diphoton rate enhancement,
the unitarity violation scale comes down to around 5\,TeV.  This is
true even when $x=0$, i.e., when the gauge coupling of the Higgs boson
matches the SM value and therefore the $WW\to WW$ scattering is
perfectly unitary.

To summarize, even though the existence of a Higgs-like particle has
been announced, precise measurements of its couplings to gauge bosons
and fermions would take quite a while.  The expected precision of the
gauge and Yukawa couplings of the Higgs is unlikely to get better than
about 25\% within a year from now \cite{Incandela}.  If the measured
couplings eventually match their SM values, the theory is unitary,
i.e.  well-behaved up to arbitrarily high energies.  Otherwise, the
extent of departure of the measured values of the couplings from their
SM predictions would mark the scale where unknown dynamics would set
in (see e.g.~\cite{Bellazzini:2012tv}).  We have carried out a
quantitative study of this scale as a function of the deviation of the
Higgs couplings from their SM values through studies of the $WW \to
WW$ and $t\bar t \to WW$ scattering processes.  We have specifically
focused on nonstandard effects on the gauge coupling of the Higgs and
the top Yukawa coupling, as these two couplings play a crucial r\^ole
in the stability of the electroweak vacuum and the perturbative
unitarity of the theory.  If future measurements favor Higgs couplings
closer to its SM values, the expected scale of unitarity saturation
would go up.

{\em Note added:}
While this work was being completed, we became aware of a similar work
\cite{Choudhury:2012tk} which has addressed similar questions.

{\em Acknowledgements:} We thank Amitabha Lahiri, Debmalya
Mukhopadhyay and Dilip K.\ Ghosh for useful discussions.  DD thanks
Department of Atomic Energy for financial support.

\end{document}